\documentclass[twocolumn,superscriptaddress,secnumarabic,amssymb,aps,prl,nobibnotes,numerical]{revtex4-1}

\usepackage{graphicx}
\usepackage[squaren]{SIunits}
\usepackage{verbatim}
\usepackage{mathrsfs}
\usepackage{braket}
\usepackage{amsmath}

\usepackage[normalem]{ulem}
\usepackage{color}
\newcommand{\be} {\begin{equation}}
\newcommand{\ee} {\end{equation}}

\begin{document}

\title{Self-protected polariton states in photonic quantum metamaterials}


\author{Matteo Biondi}
\address{Institute for Theoretical Physics, ETH Zurich, 8093 Z\"urich, Switzerland}

\author{Sebastian Schmidt}
\address{Institute for Theoretical Physics, ETH Zurich, 8093 Z\"urich, Switzerland}

\author{Gianni Blatter}
\address{Institute for Theoretical Physics, ETH Zurich, 8093 Z\"urich, Switzerland}

\author{Hakan E.\ T\"ureci}
\address{Department of Electrical Engineering, Princeton University, 
08544 Princeton, New Jersey, USA}
\address{Institute for Quantum Electronics, ETH Zurich, 8093 Z\"urich, 
Switzerland}

\begin{abstract}
We investigate the single-photon transport properties of a one-dimensional
coupled cavity array (CCA) containing a single qubit in its central site by
coupling the CCA to two transmission lines supporting propagating bosonic
modes with linear dispersion.  We find that even in the nominally weak
light-matter coupling regime, the transmission through a long array exhibits
two ultra-narrow resonances corresponding to long-lived self-protected
polaritonic states localized around the site containing the qubit. The
lifetime of these states is found to increase exponentially with the number of
array sites in sharp distinction to the polaritonic Bloch modes of the cavity
array.
\end{abstract}
\maketitle

Cavity QED studies the non-equilibrium dynamics of quantum emitters (atoms or
qubits) coupled to discrete photon modes of an electromagnetic resonator.
Such systems are of importance in the study of fundamental properties of open
quantum systems as well as for quantum information processing applications. Of
particular inte\-rest is the strong-coupling regime of cavity QED achieved when
a single excitation can be coherently exchanged between the qubit and a 
single photon mode before leaving the cavity. This regime is characterized by
well-defined quasi-particles, loosely referred to as polaritons, that are
adjustable mixtures of photonic and material excitations.\\ In recent years,
we are witnessing the shift of focus towards larger cavity QED architectures
which display collective effects due to the interaction of many photonic modes
and qubits. Such collective effects can arise in a multitude of ways. 
Complex states of matter, such as atomic \cite{baumann_dicke_2010,
mottl_rotontype_2012, ritsch_cold_2013} and polaritonic
\cite{deng_excitonpolariton_2010, carusotto_quantum_2013} condensates
represent examples where many emitters are coupled coherently to a single
cavity. Waveguide QED
\cite{shenfan_20051ph_scqubit,Chang_spt2007,ZhNori_1photscatt_CCAq,Rakh_PMM_2008,
Zheng_fps_tls_2010, Roy_stps_wg, Cicc_cca, Longo_12,
Plet_smp_oned_2012,john_bgap_bs,LSZ_shisun,Longo_1p2011_PRA, astafiev_wgqed_2010,Lang_hm_2013} and
photonic impurity systems  \cite{le_hur_kondo_2012,goldstein_inelastic_2013}
constitute examples, where single or few emitters interact with many optical
modes.  Recent theoretical work on cavity QED lattices
\cite{hartmann_quantum_2008, tomadin_manybody_2010, houck_chip_2012,
schmidt_circuit_2012} addressed systems that display interactions of
many emitters with many modes.

\begin{figure}[b]
\centering
\includegraphics[scale=0.68]{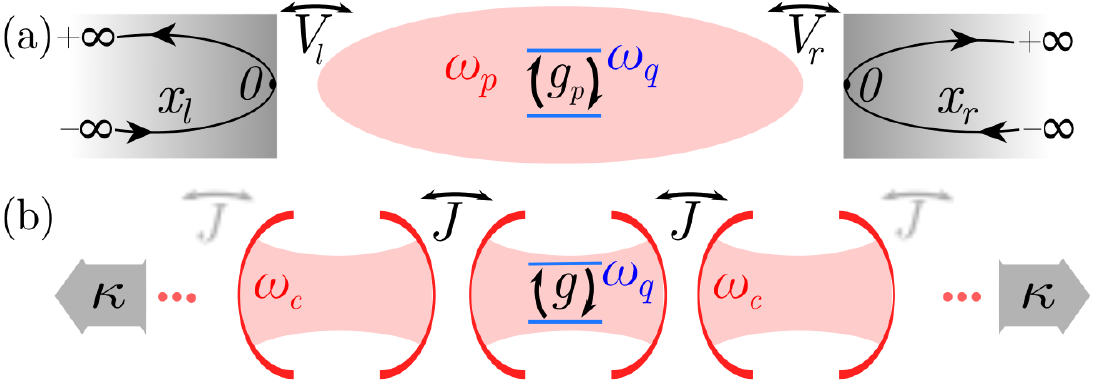}
\caption{(Color online). Scheme of a general quantum metamaterial (MM) with
photonic modes $\omega_p$ (red) coupled to a qubit of frequency $\omega_q$
(blue) with strengths $g_p$ (a). The MM is coupled to two wave\-guides with
branch coordinates $x_{l,r}$. The case $x_{l,r} > 0$ describes left (right)
propagating modes with group velocity $v_g$ in the left (right) waveguide,
while for $x_{l,r} < 0$ the propagation direction is reversed. Both
wave\-guides couple to the MM at $x_{l,r} = 0$ via the photon hopping strength
$V_{l,r}$. Panel (b) shows the specific MM considered in this paper, an array
of $N$ coupled cavities with frequency $\omega_c$, photon hopping strength $J$
and a qubit in the central cavity with coupling strength $g$. The array
couples to the wave\-guides via the outer cavities leading to a
characte\-ristic photon escape rate $\kappa = |V_{l,r}|^2/v_g$.
\label{fig1_qm_total_wg}}
\end{figure}

In this paper, we investigate the strong-coupling physics and polariton
formation when a single qubit is coupled to an open mesosopic photonic system.
From the perspective of the qubit, the mesoscopic system represents a
dissipative electromagnetic environment featuring a spectral density that is
highly structured. In the absence of additional qubit decay channels,
long-lived polaritonic states typically are formed in such a system when the
rate of coherent exchange between the qubit and the photonic modes is much
faster than the photonic decay rate $\kappa$. Here, we show that light-matter
interaction in a mesoscopic environment can result in polaritonic
modes with a hugely enhanced lifetime.  For this we consider a one dimensional
array of $N$ coupled cavities containing a single qubit in its central site
(see Fig.\ \ref{fig1_qm_total_wg}). This photonic quantum metamaterial (MM) \cite{Quach_PMM_2011,Macha_PMM_2013} is coupled to wave\-guides that constitute the sole channel for dissipation and
allow us to probe the system through photon scattering. Such a setup can be
realized in circuit QED platforms with existing technology
\cite{ houck_chip_2012,
schmidt_circuit_2012, underw_pra2012, QIproc_lucmart_2012}. Even in the regime of nominally weak light-matter
coupling, we find, among other broader features, two transmission peaks of
nearly vanishing linewidth, corresponding to \emph{quasi-bound} photon-qubit
states localized around the central cavity.  These are finite-lifetime modes
that derive from photon-qubit bound states when the cavity array extends to
infinity \cite{john_bgap_bs, LSZ_shisun, Longo_1p2011_PRA}. As
long as the hopping rate $J$ between cavities is smaller than the qubit-cavity
coupling $g$, the lifetime $\tau$ of these modes is shown to scale
exponentially with the number $N$ of array sites, $\tau \sim
(g/J)^{N-1}/\kappa$, in clear distinction with the other polaritonic Bloch
modes of the array. This result demonstrates the existence of
interaction-induced self-protected polariton states of different origin than
the remaining photon-dominated Bloch modes of the lattice.

The combined system of quantum metamaterial (MM) and waveguide is described by
the Hamiltonian $H = H_{\rm \scriptscriptstyle MM} + H_{w} + H_{{\rm
\scriptscriptstyle MM}w}$, where the first term describes the metamaterial in
terms of a finite set of local bosonic- and qubit operators, $a_j$,
$j=1,\dots,N$ and $\sigma_i^\pm$, $i=1,\dots,K$.  Here, we consider as a
metamaterial an array of $N$ coupled-cavities ($N$ odd) with frequency
$\omega_c$ containing one ($K=1$) resonant two-level system with $\omega_q =
\omega_c$ at its middle site $j=s$ [see Fig.\ \ref{fig1_qm_total_wg}(b); we
set $\hbar =1$],
\be
   \label{h_ccaq}
   H_{\rm \scriptscriptstyle MM} = H_{\rm \scriptscriptstyle CCA}
   + \omega_q\,\sigma^+\sigma^- 
   + g\,(\sigma^+a_s +a_s^\dagger\sigma^-)
\ee
with
\be
    H_{\rm \scriptscriptstyle CCA} = \omega_c \sum_{j=1}^N\,a^\dagger_j\,a_j 
   - J\sum_{j=1}^{N-1} (a_j^\dagger\,a_{j+1} + \text{hc}).
\label{h_cca}
\ee
The second term in $H_{\rm \scriptscriptstyle CCA}$ describes photon hopping
between nearest neighbours at a rate $J$, with a bare photon bandwidth
$4J\cos[\pi/(N+1)]\approx 4J-{\cal O}(J/N^2)$.  The qubit couples to the
cavity photon at site $s$ with strength $g$.  The other terms in $H$ describe
the kinetic energy of the wave\-guides
\be
   H_{w} = -iv_g \sum_{\alpha=l,r} \int_{-\infty}^{+\infty}  \!\!\!\!\!\!\!
   dx_\alpha\,\Psi_\alpha^\dagger (x_\alpha)\,
   \partial_{x_\alpha}\Psi_\alpha(x_\alpha)
   \label{h_wg_chiral}
\ee
and the metamaterial-waveguide coupling
\be
   H_{{\rm \scriptscriptstyle MM}w} 
   = \sum_{\alpha = l,r} \int_{-\infty}^{+\infty} \!\!\!\!\!\!\!
   dx_\alpha\,\delta(x_\alpha)\,V_\alpha\Psi_\alpha (x_\alpha)
   \,a^\dagger_\alpha +\text{hc}.
\ee
Here, the operator $\Psi_\alpha (x_\alpha)$ destroys a photon with group
velocity $v_g$ in the left ($l$) or right ($r$) waveguide at the branch
coordinate $x_\alpha$ [see Fig.\ \ref{fig1_qm_total_wg}(a)]. $H_{{\rm
\scriptscriptstyle MM}w}$ describes photon hopping with coupling constants
$V_\alpha$ between the CCA boundary sites and the waveguides (we defined
$a_l\equiv a_1$ and $a_r \equiv a_N$). The linear dispersion model \eqref{h_wg_chiral} 
is well suited for wave\-guides supporting TEM modes, e.g., in circuit QED \cite{wallraff_nat2004}. 
A fundamental mode with linear dispersion is found also in other bosonic
wave-guiding systems, e.g., surface plasmons in metallic nanowires
\cite{cheng_plasmnw_qd_2010} and photonic crystals wave\-guides
\cite{joann_pc}.
\begin{figure}[t]
\centering
   \includegraphics[width=0.415\textwidth]{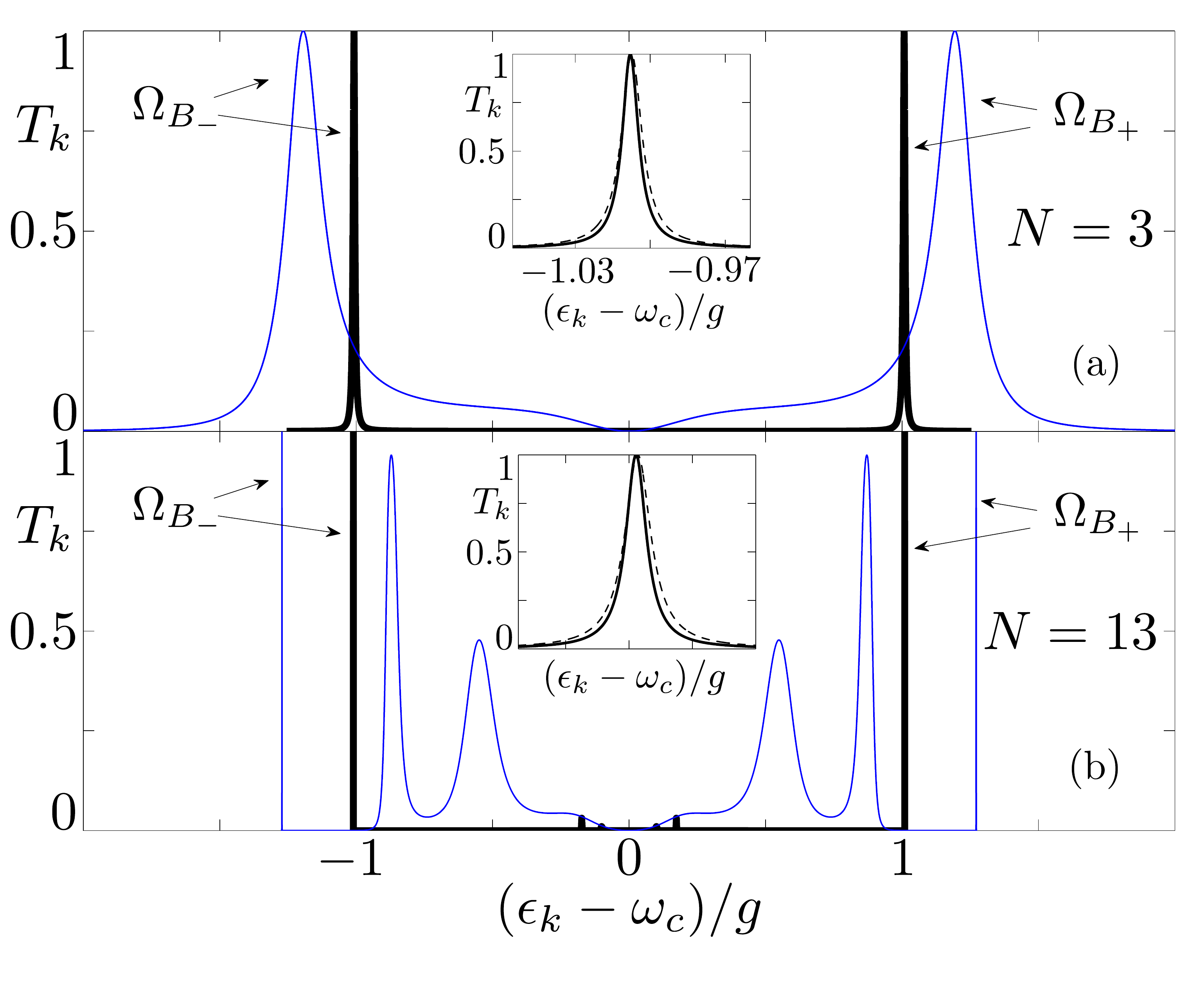}
\caption{(Color online). Transmission $T_k =|t_k|^2$ as a function of the photon energy
$\epsilon_k$ for an array with $N = 3$ (a) and $N = 13$ (b) cavities
containing a qubit in the central site, with hopping rate $J/g = 0.1$ (black, thick)
and $J/g = 0.5$ (blue, thin) for $\kappa = g$. The ultra-narrow resonances denoted
by $\Omega_{B_\pm}$ are associated with polaritonic quasi-bound states localized
around the center site of the cavity-array (see text). The insets show a zoom
into the left ultra-narrow resonance for $J/g=0.1$. Dashed lines correspond to
the analytic approximation in Eqs.\ \eqref{Tcoeff_sk} and
\eqref{linewidth_BS_general_sJ}. The other peaks ($N=13$, blue and thin) are associated
with overlapping photon-like resonances (hardly visible at $J/g = 0.1$).
\label{fig_bs1}}
\end{figure}

\begin{figure}[t]
\centering
    \includegraphics[width=.415\textwidth]{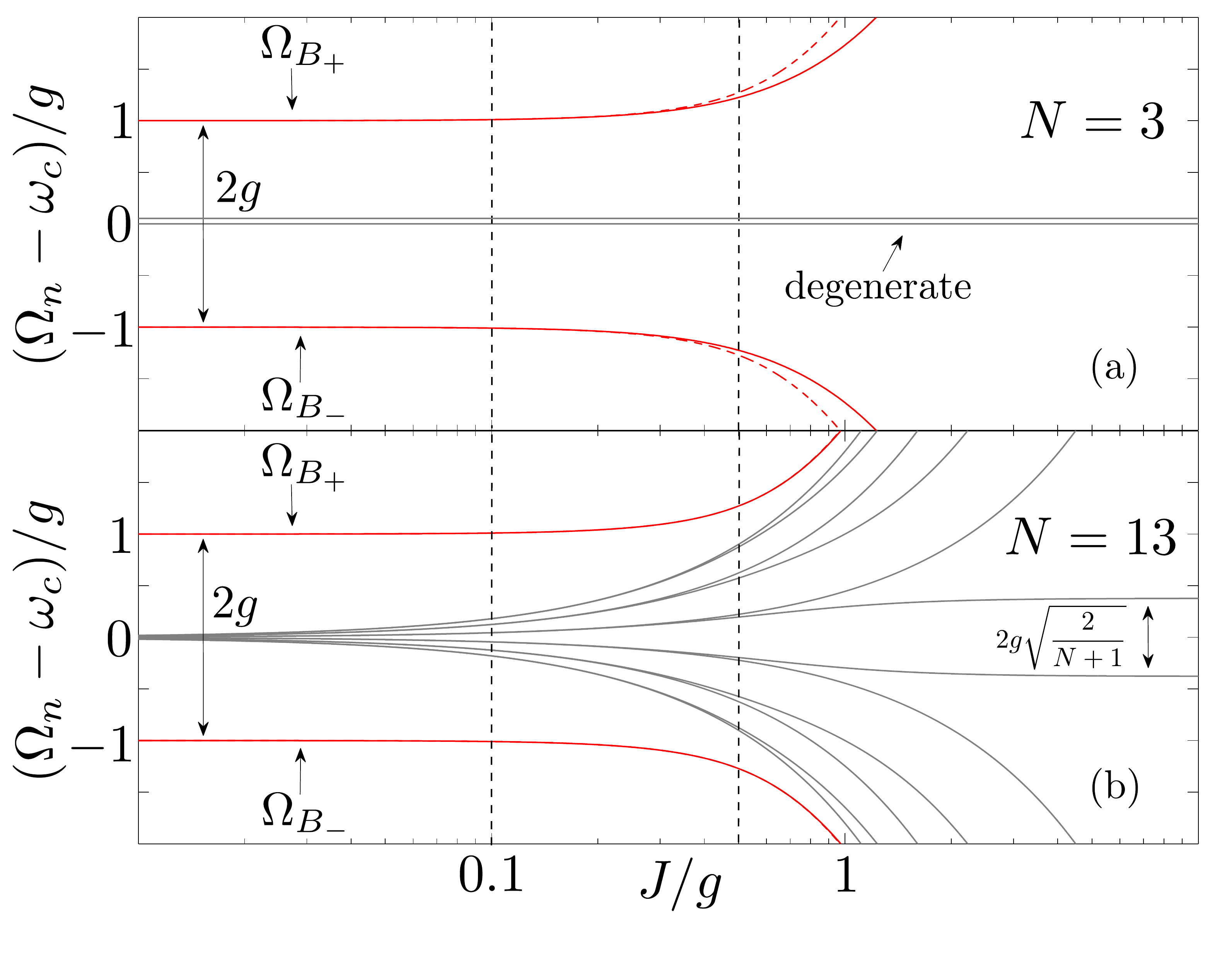}
\caption{(Color online). Eigenenergies of the quantum metamaterial in
Eq.\ \eqref{h_ccaq} as obtained from a numerical solution of
Eq.\ \eqref{Eigeneq_Omega} with $N = 3$ (a) and $N =13$ (b) cavities
plotted as a function of the effective hopping strength $J/g$. The red lines
depict the energies of the out-of-band states giving rise to the ultra-narrow
resonances seen in Fig.\ \ref{fig_bs1} (the vertical dashed lines mark the
values of $J/g$ considered there).  The red dashed lines correspond to the
analytic approximation in Eq.\ \eqref{Omega_bs_hwbcM} valid for a large number
of cavities $N \gg 1$.  \label{fig_bs2}} \end{figure}

In order to calculate the transmission, it is convenient to write the
Hamiltonian of the MM in its eigenbasis, i.e., $H_{\rm
\scriptscriptstyle MM}\,$=$\,\sum_n
\Omega_nP^\dagger_nP_n$ with eigenvalues $\Omega_n$ and the projectors
$P_n=\ket{\mathrm{vac}}\bra{n}$, which destroy an excitation in the
eigenstates $\ket{n}$. Since the total Hamiltonian $H$ commutes with the
excitation operator $N_{\mathrm{ex}}=\sum_{\alpha=l,r} \int_{-\infty}^{+\infty}
dx_\alpha\,\Psi_\alpha^\dagger(x_\alpha) \Psi_\alpha(x_\alpha)+\sum_n
P^\dagger_nP_n$, we can compose the one-excitation ansatz for the
combined metamaterial-waveguide system in the form
\be\nonumber
   \ket{\phi_k} = \bigg[\sum_{\alpha=l,r}\int_{-\infty}^{+\infty} \!\!\!\!\!\!\!
   dx_\alpha\,\phi^{\alpha}_k(x_\alpha)\Psi_\alpha^\dagger (x_\alpha) 
   +\sum_n p_k^n P^\dagger_n\bigg]\ket{\mathrm{vac}}.
\ee
Solving the Schr\"odinger equation, we obtain the Lipp\-mann-Schwinger (LS)
states for a MM of length $2L$
\be\begin{split}
   \phi^{l}_k(x_l) & = e^{ik(x_l-L)}\,[\,\theta(-x_l) + r_k\,\theta(x_l)\,],\\
   \phi^{r}_k(x_r) & = e^{ik(x_r-L)}\,t_k\,\theta(x_r),
   \label{sc_ansatz}
\end{split}
\ee
with energies $\epsilon_k = v_g k$ and the transmission and reflection amplitudes
\be
   \label{t_ampl_general}
   t_k  = \frac{-2i\beta}{\Gamma_l \Gamma_r + |\beta|^2}\quad\mbox{and}\quad
   r_k  = \frac{\Gamma_l^* \Gamma_r - |\beta|^2}{\Gamma_l \Gamma_r + |\beta|^2},
\ee
where
\be
   \Gamma_{l,r} = 1+\frac{i}{2v_g}\sum_n\,\frac{|V_n^{l,r}|^2}{\epsilon_k - \Omega_n},\>\>
   \beta =  \frac{1}{2v_g}\sum_n\,\frac{V_n^{l}\,[V_n^{r}]^*}{\epsilon_k - \Omega_n}.
\label{en_dep_cc_wgjchm}
\ee
The couplings $V_n^{l,r}=V_{l,r}\, u^{1,N}_n $ involve the photonic amplitudes
$u^j_n\,$=$\,\braket{\mathrm{vac}| a_j |n}$ at the edges.  The probability
amplitude for the excitation of the mode $n$ in the metamaterial associated with 
the LS state $k$ is given by
\be
   \label{eq_pn}
   p_k^n  = \frac{e^{-ikL}}{2(\epsilon_k - \Omega_n)}
   \big[V_n^{l} (1+ r_k) + V_n^{r} t_k].
\ee
The results Eqs.\ \eqref{t_ampl_general} -- \eqref{eq_pn} provide the general
solution of the one-photon scattering problem for any kind of MM which is
connected to two transmission lines; the specific properties of the MM at hand
are encoded in the energies $\Omega_n$ and amplitudes $u^{1,N}_n$. For the MM
in Eq.\ \eqref{h_ccaq} the eigenvalues $\Omega_n$ and eigenstates $\ket{n}$
are calculated numerically. In Fig.\ \ref{fig_bs1} we show the transmission
$T_k=|t_k|^2$ for a CCA with $N=3,13$ cavities; given the coupling $g$, we
choose $V_{l,r} =V = \sqrt{v_g g}$ generating a `weak' coupling situation
$\kappa = V^2/v_g = g$.  In both cases we find two well-resolved peaks with
ultra-narrow linewidths. As we will explain below, these peaks describe high-Q
quasi-bound polaritonic states localized around the central cavi\-ty. The
linewidth is smaller for weaker hopping strength $J$ (black, thick curves) and larger
array size $N$ [Fig.\ \ref{fig_bs1}(b)]. Bound states \cite{Sols_modtransist_1989} and the effect of quasi-bound states on transmission \cite{shao_twgres_1994} were also found in mesoscopic electronic systems, specifically in the context of ballistic transport through narrow wires with stubs or side-coupled dots. Let us then analyze the properties of the quasi-bound polaritonic states for the present situation of a photonic metamaterial. To this end, we first find the eigenvalues and eigenfunctions of the metamaterial
Eq.\ \eqref{h_ccaq}. Solving $H_{\rm\scriptscriptstyle MM} \ket{n}=
\Omega_n\ket{n}$ for a single excitation, one obtains a
non-linear equation for the eigenvalues $\Omega_n$, 
\be
   \frac{g^2}{\Omega_n - \omega_q}\,\sum_{p=1}^N 
   \frac{|\alpha^s_{p}|^2}{\Omega_n - \omega_p} = 1,
   \label{Eigeneq_Omega}
\ee
with $\omega_p=\omega_c-2J\cos[\pi p/(N+1)]$ the energies of the bare CCA and
the photonic amplitudes $\alpha^s_{p}=\sqrt{2/(N+1)} \sin[\pi p s/(N+1)]$
evaluated at the site $s$ of the qubit.  When $|\Omega_n - \omega_c|<2J$,
there are $N-1$ \emph{in-band} solutions, while for $|\Omega_n -
\omega_c|>2J$, the sum in \eqref{Eigeneq_Omega} can be evaluated analytically
for $N\gg 1$ and one obtains two out-of-band states with energies
\be
\label{Omega_bs_hwbcM}
   \Omega_{B_\pm} = \omega_c \pm \sqrt{2J^2 + \Omega^2}\quad\mbox{with}
   \quad \Omega^2=\sqrt{4J^4 + g^4}.
\ee
The exact energies as obtained numerically from Eq.\ \eqref{Eigeneq_Omega} are
shown in Fig.\ \ref{fig_bs2} together with the analytic result in Eq.\
\eqref{Omega_bs_hwbcM} (dashed lines). We observe that the bound-state
energies $\Omega_{B_\pm}$ agree with the position of the ultra-narrow
resonances in Fig.\ \ref{fig_bs1} up to a small dissipation-induced shift
to be discussed later. The corresponding eigenfunctions describe
entangled photon-qubit states, which are localized around the central cavity.
In the weak-hopping regime ($J \ll g$) and for $N \gg 1$, we can find an
approxi\-mate form of the wave-functions by assuming periodic boundary
conditions, yielding
\be
   \ket{{B_\pm}} = \frac{1}{\cal{N}}\sum_{j=1}^N e^{-|j-s|/\xi} 
   \left( \alpha\, a^\dagger_j\pm g\, \delta_{js}\sigma_j^+\right)
   \ket{\mathrm{vac}},
   \label{BS_CCAq_rspace}
\ee
with the localization length
\be
   \label{loclength}
   \xi=  -{1}/{\ln \eta}, \qquad
   \eta = _{_+}\!\!\!\sqrt{{(\Omega^2-g^2})/{2J^2}}.
\ee
Furthermore, $\alpha = _{_+}\!\!\!\sqrt{2J^2 + \Omega^2}$ and ${\cal
N} = _{_+}\!\!\!\sqrt{f\alpha^2 +g^2}$~with $f= (1+\eta^2
-2\eta^{N+1})/(1-\eta^2)$; these results are consistent with those obtained in
Refs.\ \cite{LSZ_shisun,Longo_1p2011_PRA}, where one qubit-cavity system
inside an infinitely extended CCA with periodic boundary conditions was
considered. In order to obtain experimentally measurable signatures of these
self-protected polaritonic bound states, e.g., in circuit QED, the finite size
and the coupling of the metamaterial to an external transmission line must be
taken into account.  As we will show below, the role of the wave\-guide
continuum is important in defining the interesting properties of the bound
states \eqref{BS_CCAq_rspace} in a realistic open system.  Note, that in
circuit QED, excitation and read-out can be achieved by making use of a
\emph{single-photon source} on one- and a microwave detector on the other end
of the two transmission lines enclosing the CCA \cite{Eichler_sps}.

We now analyze the transmission close to the bound states, i.e., assuming
$\epsilon_k=\Omega_{B_\pm}+\delta$, with $|\delta|\ll\Omega_{B_\pm}$. In the
weak hopping regime ($J \ll g$), the transmission can be cast into a
Lorentzian form
\be
   T_k \approx \frac{W_{B_\pm}^2}{[\delta \pm (\kappa/2g)W_{B_\pm}]^2 
   + W_{B_\pm}^2}\label{Tcoeff_sk},
\ee
describing the transformation of the bound states \eqref{BS_CCAq_rspace} into
quasi-bound states with a lifetime $\tau_{B_\pm}= 1/W_{B_\pm}$ and a frequency
shift $\mp (\kappa/2g)W_{B_{\pm}}$ due to dissipation (we have dropped terms
$\propto \kappa^2/g^2$). The Lorentzian line shapes compare well with the
exact numerical result as can be seen in the insets of Fig.\ \ref{fig_bs1}. A central result of this work then is the narrow linewidth 
\be
   W_{B_{\pm}} \approx \frac{\kappa}{2}\,\left(J/g\right)^{N-1} 
   + \kappa\,\mathcal{O}\left[\left(J/g\right)^{N+1}\right].
   \label{linewidth_BS_general_sJ}
\ee
decreasing exponentially in $N$ within the weak-hopping regime $J \ll g$; the
narrow peaks describe two long-lived, self-protected polariton resonances that
are formed by a qubit-induced localization of a photon. It turns out, that,
within a circuit QED setup, an array of $N=3$ cavities is already sufficient to support a
long-lived polariton mode around its middle site.  Such a three-resonator
device is an architecture that is readily realizable today
\cite{houck_chip_2012,
schmidt_circuit_2012, underw_pra2012, QIproc_lucmart_2012}; in these experiments, 
typical values for the lead coupling
are $\kappa = 10~\kilo \hertz - 80~\mega\hertz$, while the
light-matter coupling $g$ and photon tunneling $J$ go up to a few $100~\mega\hertz$. 
The cavity- and qubit-decay rates are in the $\kilo\hertz$
regime and thus small in comparison with $g$, $J$, and $\kappa$
\cite{schmidt_circuit_2012}; in our analysis, we have neglected such
additional dissipative processes.  Choo\-sing $\kappa = 80~\mega\hertz$, $g =
200~\mega\hertz$, $N=3$ and varying $J$ between $(1 - 100)~\mega\hertz$,
the quasi-bound states' lifetime ranges between $10^{-7}~\sec \le \tau_{B_\pm} \le
10^{-3}~\sec$, much larger than the original bare lifetime $\kappa^{-1} \sim
10^{-8}~\sec$.

\begin{figure}
\centering
\includegraphics[width=.415\textwidth]{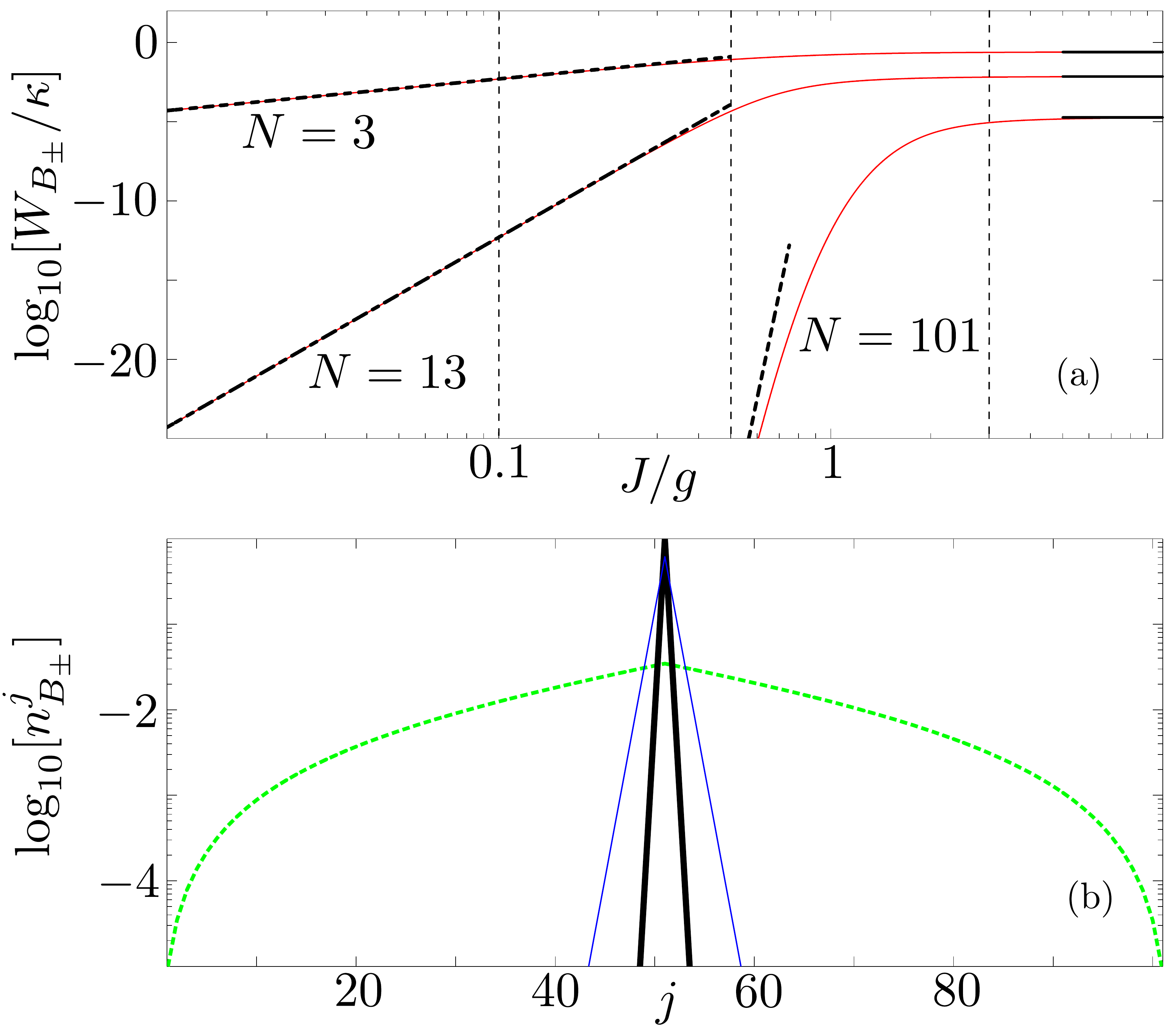}
\caption{(Color online). Panel (a) shows the exact linewidth of the resonances at
$\Omega_{B_\pm}$, see Fig.\ \ref{fig_bs1}, as a function of hopping strength
for $N=3,\,13,\,101$ cavities. The black dashed (solid) lines correspond to the
asymptotic results for weak (large) photon hopping as described in Eqs.\
\eqref{linewidth_BS_general_sJ} and \eqref{linewidth_BS_general_lJ}.  Panel
(b) shows the photonic occupation probability $n^j_{B_\pm}$ with site index $j$, 
evaluated at the quasi-bound state energies $\epsilon_k\approx \Omega_{B_\pm}$, for 
$N=101$ cavities and $J/g=0.1$ (black, thick), $J/g=0.5$ (blue, thin) and $J/g=3$ 
(green, dashed). \label{fig_bs3}}
\end{figure}

Increasing the hopping $J \gg g$, the polariton states transform into photonic
states which are again described by a Lorentzian but with a linewidth that
decreases only algebraically with $N$,
\be
   W_{B_\pm} = \frac{2\kappa}{N+1}\sin^2[\pi/(N+1)] \sim 
   \frac{2\pi^2\,\kappa}{N^3}.
   \label{linewidth_BS_general_lJ}
\ee
The exact linewidth $W_{B_{\pm}}$ is plotted in Fig.\ \ref{fig_bs3}(a) as a
function of the effective hopping strength $J/g$ together with the
approximations at weak- \eqref{linewidth_BS_general_sJ} and large hopping
\eqref{linewidth_BS_general_lJ}.  In Fig.\ \ref{fig_bs3}(b) we plot the
photon occupation probability in the MM evaluated at the quasi-bound state
energies, i.e., $n^j_{B_\pm} =\braket{\phi_k| a^\dagger_j a_j
|\phi_k}|_{\epsilon_k \approx \Omega_{B_\pm}}$, for an array with $101$
cavi\-ties.  At weak hopping $J \ll g$, the excitation is loca\-lized in a
narrow region around the central cavity (black, blue lines) while it
delocalizes over the lattice at large hopping $J \gg g$.

Finally, we comment on the in-band resonances, e.g., the broad peaks appearing
in Fig.\ \ref{fig_bs1}(b) at $J/g = 0.5$. In order to do so, it is helpful to
analyze the spectra in Fig.\ \ref{fig_bs2}. Given the photonic and qubit
degrees of freedom, we expect a total of $N+1$ states of which $N-1$ should
be located inside the band.  At small $J$, the central cavity mixes with the
qubit to generate the (quasi-)bound states at $\omega_c \pm g$; for $N=3$ no
oscillator-strength at the central cavity is left for the other two modes
(this remains true for all $J$) and we find a two-fold degenerate state at
$\Omega_n = \omega_c = \omega_q$ describing a pure photon and a photon-like
state with small qubit-weight which does not mix. Increasing $J$, the pure
photon state remains, the photon-dominated state transforms into a
qubit-dominated one, and the (quasi-)bound states become photon-dominated.  In
general, for odd $M=(N-1)/2$, the spectrum still features a doublet of
degenerate states at $\omega_c$ with the same properties as described above
for $N=3$; however, for $N>3$, some weight of the central cavity is
transferred from the quasi-bound states to the other in-band modes with
increasing $J$.  For even $M$, the two states near $\omega_c$ are both
photon-like at small $J$ and their mi\-xing with the qubit with increasing $J$
results in a splitting $2g\sqrt{2/(N+1)}$.  At small $J$, i.e., to order $J$,
the re\-sonances within the band appear in pairs, reducing the number of
expected peaks by half.  Finally, the transmission is exactly zero right at
the qubit frequency due to the so-called dipole-induced reflectivity effect
(DIR)\cite{DIT_vuc} known from the single-cavity case. The observed (peak)
structure in Fig.\ \ref{fig_bs1} is consistent with these considerations.

In summary, we have studied the single-photon transport properties of a
one-dimensional coupled cavity array containing a single qubit in its center,
and coupled to a transmission line supporting propagating photon modes with
linear dispersion. For small hopping $J$, the transmission coefficient
exhibits two high-Q polariton resonances associated with self-protected
quasi-bound states of the qubit and photonic Bloch modes. We have derived a
simple expression for the lifetime of these states and found that it increases
exponentially with the number of sites, thereby realizing a regime of strong
light-matter coupling. The proposed architecture is readily realizable within
current state-of-the-art technology, e.g., in circuit QED
\cite{ houck_chip_2012,
schmidt_circuit_2012, underw_pra2012, QIproc_lucmart_2012}.

We acknowledge financial support from the Swiss NSF through an Ambizione Fellowship (SS) and under Grant No.\ PP00P2-123519/1, the NCCR QSIT (MB) and the National Science Foundation under Grant No.\ DMR-1151810.

\bibliography{references_CCAPOL}

\end{document}